\documentclass[a4paper,twoside]{article} 

\input{obsmag.sty}   

\def\t0{$T_0$}

\OBSheader{The Eruption of M31N 2017-9a}{C. Lloyd et al.}{2020 June}{140}

\begin{document}

\OBStitle{The Unusual Eruption of the Extragalactic \\ 
	Classical Nova M31N 2017-09a}

\OBSauth{
	Christopher Lloyd\,$^1$, 
	Lewis M. Cook\,$^{2,3}$, 
	Seiichiro Kiyota\,$^{4,5}$, and\\
	Hanjie Tan\,$^{6,7}$,
	Di Hu\,$^7$,
	Wei Shi\,$^7$,
	Mi Zhang\,$^7$,
	Shenwei Zhang\,$^7$,\\
	Xiaowei Zhang\,$^7$,
	Jingyuan Zhao\,$^7$,
	Guoyou Sun\,$^7$,
	Xing Gao\,$^{7,8}$,\\
	-- Xingming Observatory Sky Survey Group (XOSS) --\\
	and David Boyd\,$^{9,10}$
}


\OBSinst{School of Mathematical and Physical Sciences, University of Sussex}

\OBSinst{CBA Concord Observatory, Concord, CA 94518, USA}
\OBSinst{American Association of Variable Star Observers}

\OBSinst{7-1 Kitahatsutomi, Kamagaya Chiba 273-0126 Japan}
\OBSinst{Variable Star Observers League of Japan}

\OBSinst{National Central University, 32001 Taiwan, Republic of China}
\OBSinst{Xingming Observatory, Mount Nanshan, Xinjiang, China}	

\OBSinst{Xinjiang Astronomical Observatory, Chinese Academy of Sciences}

\OBSinst{West Challow Observatory, Wantage, OX12 9TX, UK}
\OBSinst{British Astronomical Association, Variable Star Section}

\OBSabstract{
	M31N 2017-09a is a classical nova and was observed for some 160 days following its initial eruption, during which time it underwent a number of bright secondary outbursts. The light-curve is characterized by continual variation with excursions of at least 0.5 magnitudes on a daily time-scale. The lower envelope of the eruption suggests that a single power-law can describe the decline rate. The eruption is relatively long with $t_2 = 111$, and $t_3 = 153$ days.
}

\section*{Introduction}

A nova eruption is generally understood to be caused by a thermonuclear runaway on the surface of a white dwarf after accreting sufficient material from a cool, low-mass companion. See the reviews by
Warner \cite{1995CAS....28.....W} 
and Bode \& Evans \cite{2008clno.book.....B}, 
and the more recent paper by Starrfield \etal \cite{2016PASP..128e1001S} 
The recurrence time-scale of classical novae is long, generally $> 1000$ years
(see Shara \etal, \cite{2012ApJ...756..107S} 
and Mr{\'o}z \etal \cite{2016Natur.537..649M} 
for example)
and is related to the mass of the white dwarf.
As material accumulates on the white dwarf the recurrence interval of the eruptions shortens to $< 100$ years and the system morphs seamlessly into a recurrent nova, of which there are currently only ten galactic examples
\cite{2013IAUS..281..154A, 
2015AcPPP...2..246M}. 
The system is then on an inexorable path to a Type-1a supernova.

Novae in M\,31 are apparently easy to discover, relative to Galactic novae at least. Over the past five years 37 novae have been discovered on average per year in M\,31, which is typically 4 times the galactic discovery rate
\cite{MPEnovavM31,kojinovae}.
Of course these tend to be faint with the average discovery magnitude $V \sim 18$, so consequently few have more than the peak brightness and initial decline rate. However, some important objects have been discovered and followed up, and the most significant in recent years is the recurrent nova M31N 2008-12a discussed by 
Darnley \etal,
\cite{2014A&A...563L...9D, 
	2017ApJ...849...96D} 
which has a remarkably short recurrence interval of approximately one year.

The nova discussed here is M31N~2017-09a (AT 2017glc, PNV J00440872
+4143367) which was discovered by the XOSS Group
\cite{2017TNSTR.954....1Z} 
while making follow-up observations of M31N 2008-12a, 
and lies some 18 arc minutes distant from it.
The nova was discovered at an unfiltered (calibrated as $V$) $CV$ magnitude of 19.5 on 2017 August 30.7 UT ($JD=2457997.2$) and remained at $CV \sim 18$ for the following three days before peaking at $CV = 17.6$ then fading very rapidly to $CV \sim 20$. 
Some ten days after maximum it was confirmed spectroscopically as an Fe\,\textsc{ii} class nova by Williams \& Darnley
\cite{2017ATel10741....1W}.
Although the nova continued to be observed it did not attract any particular attention until it underwent a significant rebrightening about 100 days after discovery when it peaked at $CV = 18.3$, $R=18.2$
\cite{2017ATel11070....1T, 
2017ATel11076....1V}, 
0\fmm7 below the peak brightness.

\begin{table}[h]
	\caption{\em List of equipment used\label{equipment}} 
	\centering
	\begin{tabular}{ccc}
		{\em Observer} & {\em Telescope} & {\em CCD} \\ 
		\addlinespace[3pt]
		Boyd & LX 200 0.35-m SCT & SXVF-H9 \\
		Cook & 0.73-m reflector & STF8300M \\
		Cook & CDK24 0.61-m\,$^{\rm a}$ & PL09000  \\
		Cook & CDK20 0.51-m\,$^{\rm b}$ & PL11002MT11  \\
		Kiyota & CDK24 0.61-m $^{\rm a}$ & PL09000 \\
		XOSS & NEXT 0.60-m & PL230  \\
		XOSS & HMT 0.50-m SCT & QHY-16 \\
		XOSS & C14 0.35-m SCT & QHY-9 \\
	\end{tabular}
	
	\vspace{6pt}
	a) iTelescope T24 MPC U69 Sierra Remote Observatory, Auberry, CA\\
	b) iTelescope T11 MPC H06 New Mexico Skies, Mayhill, NM
	%
\end{table}

\section*{Observations}

The observations were made using several telescopes ranging in size from 0.35-m to 0.73-m based in the USA, China and the UK, and with several different detectors. All the images were de-biased and flat fielded by the respective observers prior to image processing. In order to improve the signal to noise it was necessary to stack a number of images depending on a variety of constraints and this varied from typically five images up 30 to in one case. The magnitude of the transient was determined relative to a comparison sequence\footnote{Sequence available through the AAVSO Variable Star Plotter} through professional or commercial aperture-photometry software. All the images bar one set were taken unfiltered, either as `Clear' or `Luminance', which is similar to clear but with an infra-red cut off. These were calibrated against a $V$-magnitude sequence to give $CV$ magnitudes. 

The early images were centred on the recurrent nova M31N 2008-12a so the sequence for that field was used, particularly the two brighter comparison stars (148 and 154) which lie between the two novae. All the observers used these comparisons except for Cook who developed what has become the AAVSO sequence for M31N 2017-09a, which is fainter and centred on this target. It is based on an extrapolation of nearby APASS standards 
\cite{2016yCat.2336....0H} 
but most of these stars are faint for APASS and have relatively large uncertainties. However, they simply provide the zero point, the sequence magnitudes are based on many measurements of the field. It also incorporates some faint stars measured by 
Massey \etal, \cite{2006AJ....131.2478M} which reach $V=22$. A comparison of the fainter sequence stars with measurements based on the brighter M31N 2008-12a (148 and 154) comparison stars shows no significant difference within the measurement errors of the nova and is small compared with its variations.

One of the stars on Massey \etal's list lies 6 arc seconds from the nova at $V=21.15$ and was clearly identified on all the faint images. For some images this star marked the limiting magnitude but on others  fainter stars could also be identified. So, despite the large uncertainties the nova was clearly identified on all the deep images.

A single set of additional $BVRI$ magnitudes was taken on one night by the XOSS team with the NEXT 0.60-m telescope. These were taken during the major rebrightening $\sim 100$ days into the eruption and give colours of $B-V = 0.39$ and $V-R = -0.05$ at that time, with uncertainties of 0\fmm1 (uncorrected for reddening). A set of five $R$-band observations reported by Valcheva \etal,
\cite{2017ATel11076....1V}
has also been used. All the observations are collected in Table~\ref{data}.

\section*{The evolution of the eruption}
 
\begin{figure}[t]
	\centering
	\includegraphics[width=0.96\textwidth]{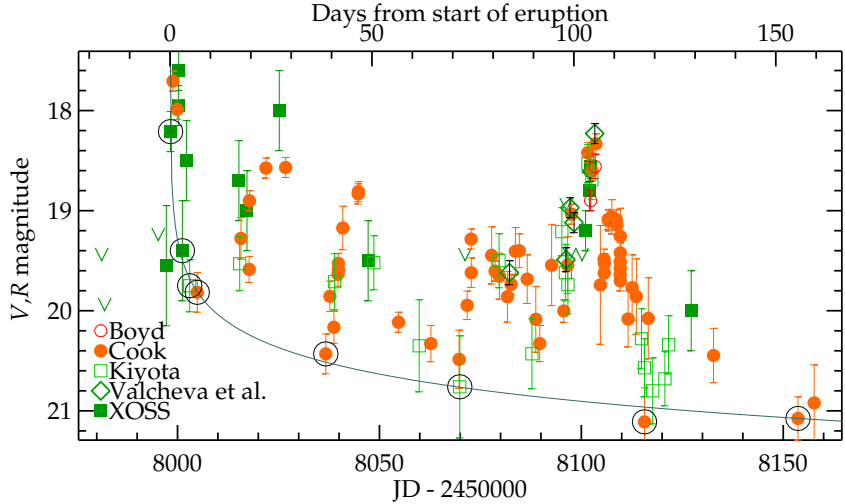} 
	\caption{Entire light-curve of the eruption. The line is the single-slope fit to the circled points which are taken to define the lower envelope of the eruption using Scenario 2 as described in the text. The single-slope fit to the first scenario is very similar, but marginally worst.}
	\label{fig:outburst}
\end{figure}

The whole eruption is shown in Fig.~\ref{fig:outburst} with the $CV$, the single $V$ and the small number of $R$ magnitudes plotted without adjustment. Given the uncertainties in the measurements and that $V \sim R$ that is not unreasonable. However, as most nova eruptions evolve the continuum cools and emission lines develop, particularly H$\alpha$ which means that ($V-R$) tends to decrease and so the $CV$ magnitudes might also be expected to brighten relative to $V$. The change in ($V-R$) could be small or up to one magnitude depending on the type of nova   
\cite{2014MNRAS.440.3402M, 
2015gacv.workE..56C, 
2015MNRAS.447.1661M, 
2016IBVS.6162....1M} 
but it is not clear how this translates into a difference between $V$ and $CV$ nor if it is even applicable to this system. In the rebrightening event ($V-R$)$\sim$~zero, and in fact all the $R$ magnitudes are consistent with contemporary $V$ or $CV$ values so any reddening of the nova must scale with magnitude rather than time, and be limited to the fainter magnitudes. Even if there is a one magnitude change in ($V-R$) if less than half of this translates to $CV$ then the fainter $CV$ magnitudes may be too bright by perhaps 0\fmm3, which is about equal to the uncertainties. So, as a working hypothesis it is assumed that for all practical purposes $CV = V = R$.

The eruption is clearly dominated by several large secondary outbursts, which although not as well observed as the documented rebrightening, are nearly as bright. These recur on an approximate time-scale of 25 days, but there is also significant short-term activity. Despite the sometimes large uncertainties there are excursions of half a magnitude or more on a daily basis. 

Despite the fragmentary nature of the data this nova is probably one of the best observed in M\,31 in recent years so every attempt has been made to derive useful parameters from it. In a major study of a large sample of novae
Strope \etal \cite{2010AJ....140...34S}
identified seven broad classes based on the light-curve shape and measured properties, while recognizing the enormous variation in the light-curves, and indeed the measured properties. For novae in Strope \etal's `smooth” class, with well behaved light-curves, there is often a well-defined power-law relationship between the magnitude and the time from the eruption, of the form, $V \propto \alpha \log(t-T_0)$ where \t0\ is time zero point of the relationship, and $\alpha$ is the slope of the power-law. Although less well defined, power-law relationships are also seen in the other classes of novae, where the power-law defines the lower envelope of the eruption. In practice well observed light-curves can show several different power-law sections with break points between them.
As well as being observed there are theoretical reasons for expecting a relationship of this kind. 
Hachisu \& Kato \cite{2006ApJS..167...59H}
have proposed a ``universal decline
law” which predicts the slope of the power-law
over time as the expanding shell evolves, with the first break point being closely related to the mass of the white dwarf.

The light-curve in Fig.~\ref{fig:outburst} seems to divide naturally into two parts, the early rapid decline and the later slow fade which is marked by large secondary outbursts. However, the light-curve shows so much short-term activity it is difficult to define any base level that marks the decline of the nova. If there are two power-law sections, then one would cover the short section at the start of the eruption with a break point after $\sim 5 - 10$ days, and the second lasting for the rest of the observed eruption. 

\renewcommand{\tabular}{\footnotesize\oldtabular}
\begin{table}[p]
	\caption{\it Photometric observations of M31N 2017–09A\label{data}} 
	\centering
	\addtolength{\tabcolsep}{-2pt}
	\begin{tabular}{crccccrccc}
		{\it JD -} & {\it Mag.} & {\it Error} & {\it Band} & {\it Observer} & {\it JD -} & {\it Mag.} & {\it Error} & {\it Band} & {\it Observer} \\
		{\it 2450000} &         &      &      &          & {\it 2450000} &         &  \\ 
		\addlinespace[3pt]
		7981.23 & $>$19.5 & 0.00 &  CV  &   XOSS   & 8095.17 &   19.21 & 0.24 &  CV  &  Kiyota  \\
		7981.96 & $>$20.0 & 0.00 &  CV  &  Kiyota  & 8095.65 &   20.00 & 0.11 &  CV  &   Cook   \\
		7995.22 & $>$19.3 & 0.00 &  CV  &   XOSS   & 8096.05 &   19.62 & 0.28 &  CV  &  Kiyota  \\
		7997.22 &   19.55 & 0.60 &  CV  &   XOSS   & 8096.21 &   19.49 & 0.12 &  R   & Valcheva \\
		7998.28 &   18.21 & 0.20 &  CV  &   XOSS   & 8096.27 & $>$19.0 & 0.00 &  CV  &   XOSS   \\
		7998.87 &   17.71 & 0.04 &  CV  &   Cook   & 8096.60 &   19.54 & 0.28 &  CV  &   Cook   \\
		7999.92 &   17.99 & 0.04 &  CV  &   Cook   & 8096.67 &   19.74 & 0.29 &  CV  &  Kiyota  \\
		8000.21 &   17.95 & 0.20 &  CV  &   XOSS   & 8097.10 & $>$19.5 & 0.00 &  CV  &   XOSS   \\
		8000.30 &   17.60 & 0.20 &  CV  &   XOSS   & 8097.19 &   18.97 & 0.09 &  R   & Valcheva \\
		8001.21 &   19.40 & 0.50 &  CV  &   XOSS   & 8097.59 &   19.04 & 0.08 &  CV  &   Cook   \\
		8002.20 &   18.50 & 0.40 &  CV  &   XOSS   & 8098.20 &   19.12 & 0.08 &  R   & Valcheva \\
		8003.02 &   19.75 & 0.26 &  CV  &  Kiyota  & 8100.16 & $>$19.5 & 0.00 &  CV  &   XOSS   \\
		8004.89 &   19.82 & 0.20 &  CV  &   Cook   & 8101.04 &   19.20 & 0.20 &  CV  &   XOSS   \\
		8015.20 &   18.70 & 0.40 &  CV  &   XOSS   & 8101.59 &   18.42 & 0.04 &  CV  &   Cook   \\
		8015.40 &   19.53 & 0.27 &  CV  &  Kiyota  & 8101.63 &   18.53 & 0.19 &  CV  &  Kiyota  \\
		8015.72 &   19.28 & 0.21 &  CV  &   Cook   & 8102.02 &   18.80 & 0.20 &  CV  &   XOSS   \\
		8017.20 &   19.00 & 0.40 &  CV  &   XOSS   & 8102.20 &   18.61 & 0.08 &  R   &   XOSS   \\
		8017.75 &   19.59 & 0.13 &  CV  &   Cook   & 8102.20 &   18.95 & 0.06 &  B   &   XOSS   \\
		8017.80 &   18.90 & 0.04 &  CV  &   Cook   & 8102.20 &   18.56 & 0.07 &  V   &   XOSS   \\
		8021.89 &   18.57 & 0.03 &  CV  &   Cook   & 8102.20 &   19.13 & 0.14 &  I   &   XOSS   \\
		8025.20 &   18.00 & 0.40 &  CV  &   XOSS   & 8102.31 &   18.90 & 0.09 &  CV  &   Boyd   \\
		8026.78 &   18.57 & 0.03 &  CV  &   Cook   & 8102.59 &   18.69 & 0.20 &  CV  &  Kiyota  \\
		8036.73 &   20.43 & 0.20 &  CV  &   Cook   & 8102.65 &   18.60 & 0.06 &  CV  &   Cook   \\
		8037.72 &   19.86 & 0.15 &  CV  &   Cook   & 8103.33 &   18.23 & 0.05 &  R   & Valcheva \\
		8038.74 &   20.17 & 0.16 &  CV  &   Cook   & 8103.42 &   18.56 & 0.12 &  CV  &   Boyd   \\
		8038.88 &   19.71 & 0.28 &  CV  &  Kiyota  & 8103.60 &   18.33 & 0.08 &  CV  &   Cook   \\
		8039.75 &   19.63 & 0.12 &  CV  &   Cook   & 8104.71 &   19.74 & 0.59 &  CV  &   Cook   \\
		8039.83 &   19.53 & 0.07 &  CV  &   Cook   & 8105.61 &   19.49 & 0.07 &  CV  &   Cook   \\
		8039.83 &   19.59 & 0.07 &  CV  &   Cook   & 8105.68 &   19.62 & 0.10 &  CV  &   Cook   \\
		8040.92 &   19.17 & 0.22 &  CV  &   Cook   & 8105.70 &   19.53 & 0.06 &  CV  &   Cook   \\
		8044.74 &   18.83 & 0.04 &  CV  &   Cook   & 8106.65 &   19.09 & 0.05 &  CV  &   Cook   \\
		8044.80 &   18.81 & 0.07 &  CV  &   Cook   & 8106.78 &   19.10 & 0.05 &  CV  &   Cook   \\
		8047.20 &   19.50 & 0.40 &  CV  &   XOSS   & 8107.54 &   19.06 & 0.17 &  CV  &   Cook   \\
		8048.61 &   19.52 & 0.27 &  CV  &  Kiyota  & 8108.63 &   19.09 & 0.04 &  CV  &   Cook   \\
		8054.73 &   20.12 & 0.07 &  CV  &   Cook   & 8108.78 &   19.14 & 0.07 &  CV  &   Cook   \\
		8059.85 &   20.35 & 0.46 &  CV  &  Kiyota  & 8109.61 &   19.64 & 0.10 &  CV  &   Cook   \\
		8062.74 &   20.33 & 0.18 &  CV  &   Cook   & 8109.63 &   19.58 & 0.10 &  CV  &   Cook   \\
		8069.73 &   20.49 & 0.29 &  CV  &   Cook   & 8109.63 &   19.42 & 0.07 &  CV  &   Cook   \\
		8069.89 &   20.76 & 0.51 &  CV  &  Kiyota  & 8109.64 &   19.51 & 0.08 &  CV  &   Cook   \\
		8071.20 & $>$19.5 & 0.00 &  CV  &   XOSS   & 8109.64 &   19.70 & 0.10 &  CV  &   Cook   \\
		8071.73 &   19.95 & 0.14 &  CV  &   Cook   & 8109.66 &   19.26 & 0.28 &  CV  &   Cook   \\
		8072.71 &   19.28 & 0.07 &  CV  &   Cook   & 8111.57 &   20.08 & 0.28 &  CV  &   Cook   \\
		8072.74 &   19.62 & 0.15 &  CV  &   Cook   & 8112.58 &   19.77 & 0.33 &  CV  &   Cook   \\
		8077.79 &   19.45 & 0.29 &  CV  &   Cook   & 8113.61 &   19.86 & 0.38 &  CV  &   Cook   \\
		8078.64 &   19.60 & 0.07 &  CV  &   Cook   & 8114.90 &   20.28 & 0.30 &  CV  &  Kiyota  \\
		8079.61 &   19.50 & 0.27 &  CV  &  Kiyota  & 8115.63 &   20.57 & 0.29 &  CV  &  Kiyota  \\
		8079.72 &   19.66 & 0.23 &  CV  &   Cook   & 8115.64 &   21.11 & 0.34 &  CV  &   Cook   \\
		8081.70 &   19.86 & 0.25 &  CV  &   Cook   & 8116.56 &   20.08 & 0.41 &  CV  &   Cook   \\
		8082.20 &   19.62 & 0.12 &  R   & Valcheva & 8117.66 &   20.80 & 0.33 &  CV  &  Kiyota  \\
		8082.60 &   19.74 & 0.09 &  CV  &   Cook   & 8120.69 &   20.68 & 0.27 &  CV  &  Kiyota  \\
		8083.67 &   19.41 & 0.24 &  CV  &   Cook   & 8121.59 &   20.34 & 0.29 &  CV  &  Kiyota  \\
		8084.63 &   19.40 & 0.17 &  CV  &   Cook   & 8127.20 &   20.00 & 0.40 &  CV  &   XOSS   \\
		8086.59 &   19.68 & 0.24 &  CV  &   Cook   & 8132.73 &   20.45 & 0.27 &  CV  &   Cook   \\
		8087.76 &   20.43 & 0.35 &  CV  &  Kiyota  & 8153.65 &   21.08 & 0.22 &  CV  &   Cook   \\
		8088.71 &   20.09 & 0.33 &  CV  &   Cook   & 8157.65 &   20.92 & 0.38 &  CV  &   Cook   \\
		8089.69 &   20.33 & 0.18 &  CV  &   Cook   & 8153.65 &   21.08 & 0.22 &  CV  &   Cook   \\
		8092.65 &   19.55 & 0.40 &  CV  &   Cook   & 8157.65 &   20.92 & 0.38 &  CV  &   Cook   \\ 
	\end{tabular}
	%
\end{table}

\begin{figure}[t]
	\centering
	\includegraphics[width=0.96\textwidth]{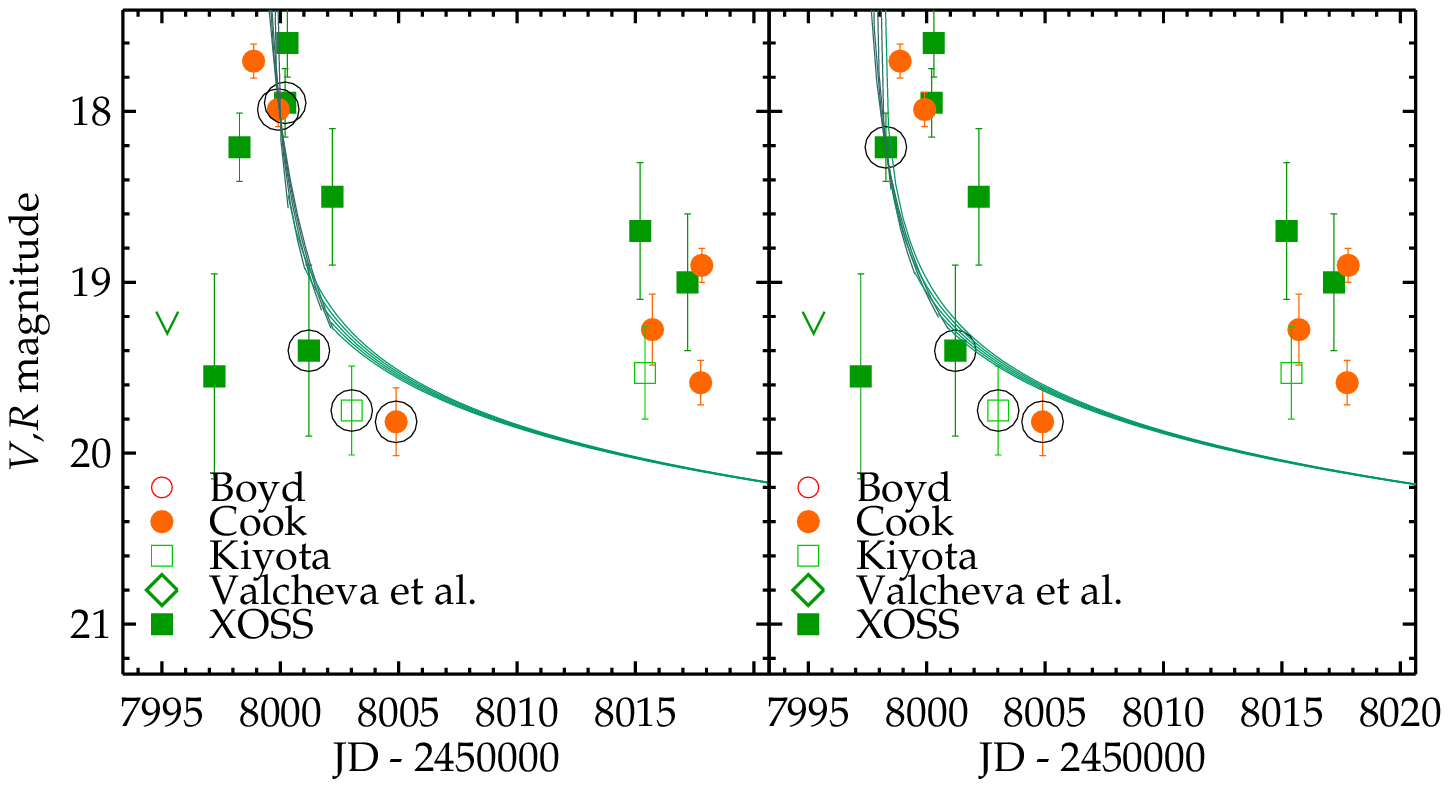} 
	\caption{The early part of the eruption showing (left) Scenario 1 the fits to the circled data for a variety of $T_0$ values as described in the text, assuming the eruption was caught on the rise.  Scenario 2 (right) similarly showing the fits to the early circled data assuming the eruption occurred between the first two positive points.}
	\label{fig:early}
\end{figure}

An attempt has been made to measure the underlying power-laws but there is so much activity even in the initial eruption that it is not completely clear how it developed.
The most obvious interpretation (Scenario 1) is that the eruption was caught on the rise and that the maximum occurred over the two days while the object was brighter than $V=18.0$. During this time there is significant variation of $\sim 0\fmm3$ and there is also the suggestion of further instability during the rapid decline to $V \sim 20$ over the following 4 days. During this interval five points have been identified that might represent the baseline of the eruption and various power-laws have been fitted assuming different values for \t0\ over one day prior to the first point.

The less obvious, alternative interpretation (Scenario 2) is that the eruption occurred earlier, in the one day between the first two positive observations, which treats all the following bright points as secondary maxima. Given the behaviour over the rest of the eruption this is not unreasonable. In this interval four points define the baseline, three of which are shared with the previous scheme, and the power-law was fitted in the same way. For the later part of the light-curve the lower envelope is very sparsely covered and only four were used as shown in Fig.~\ref{fig:outburst}. The same \t0\ values were used and the slope was fitted in the same way. The intersection of these two sets of curves defines the break point between the two power-laws, and the difference between this and \t0\ gives $t_{b1}$.

The fits to the early data are shown in Fig.~\ref{fig:early} and apart from a small offset in time show very little perceptible difference. The slope of the power-law lies in the range 2.5 -- 1.3 for Scenario 1, and  1.9 -- 1.1 for Scenario 2, and in both cases the break point falls between $\sim 0$ -- 17 days after \t0. The implication of $t_{b1}\sim 0$ is that a single power-law could fit all the data. The reason for the wide range of slopes and particularly $t_{b1}$ is that the power-law is a measure of the decline rate as the eruption unfolds. Near \t0\ it becomes unstable and tends to infinity, and this can be seen in the plots. Using all the points that define the lower envelope it is possible to fit all the parameters of the power-law using a non-linear least squares, which is not possible for the individual sections. Doing this yields $\alpha = 1.1\pm0.2$ and $0.9\pm0.2$ for Scenarios 1 and 2 respectively, and in both cases \t0\ occurs very close, and obviously prior to the first point of their respective sets. To test the effect of overestimating the brightness of the later points of the envelope they have also been set 0\fmm3 fainter, in which case the slopes are increased by 0.2.

Strope \etal\ provide useful and descriptive names for their classes of novae and at 38\% of their sample the smooth class is the most abundant. The next three most populous classes are light-curves showing plateaus (21\%), dust dips (18\%), and jitters (16\%), and it is this last class that M31N 2017-09a most closely resembles. These tend to be active for a long period before finally declining and so although the light-curve may pass 2 or even 3 magnitudes below the peak early in the eruption it is the \emph{final time} it passes this point, that determines  $t_2$, $t_3$ \etc, so it tends to be late. From Fig.~\ref{fig:outburst} it can be seen that both these points are passed very close to the end of the observed eruption giving $t_2 = 111$, and $t_3 = 153$ days. These values, particularly $t_2$, are among the largest in Strope \etal's sample.

\begin{figure}[t]
	\centering
	\includegraphics[width=.75\textwidth]{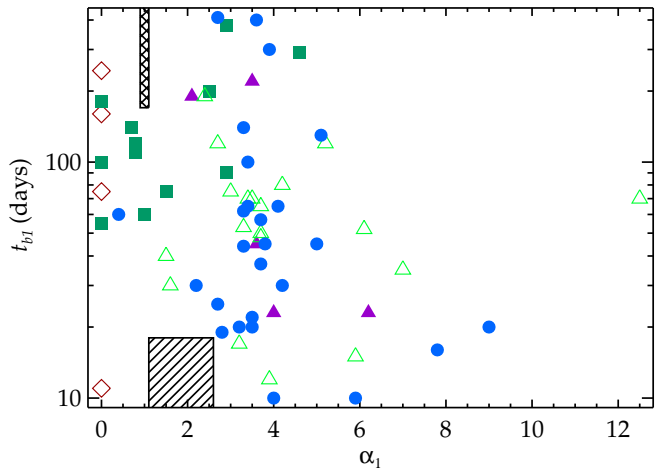} 
	\caption{Plot of the time to the first break in the power-law gradients \emph{vs.} the first gradient ($\alpha_1$) for all the novae in Strope \etal's sample for which both values exist. The main points of interest are from the `smooth' (filled circles) and the `jitter' (filled squares) light-curves. The vast majority of the `jitter' light-curves have $t_{b1} > 50$. The hatched region at the bottom of the plot covers the values of the two-slope solutions while the cross-hatched region at the top is formed by the two single-slope solutions and a lower limit of the break point from the length of the data.}
	\label{fig:tb1s1}
\end{figure}

It is also possible to compare parameters of the power-law fits with Strope \etal's sample. Fig.~\ref{fig:tb1s1}
shows the time to the first break point $t_{b1}$ \emph{vs.} $\alpha_1$, the slope of the first power-law. It should be noted that Strope \etal\ define $\alpha$ in the opposite sense ($-\alpha$) so their values have been reversed. Using  $\alpha_1$ from the two-slope fit and the limits on $t_{b1}$ gives the box at the bottom of the plot. Although the values of $\alpha_1$ are consistent with other `jitter' class novae $t_{b1}$ is not, and also the box appears to lie in a zone of avoidance. By contrast, $\alpha$ from the single power-law and a lower limit to $t_{b1}$ from the length of the data, are both consistent with similar `jitter' class novae.

\section*{Conclusions}

The eruption of M31N 2017-09a is very complex and dominated by probably four, secondary outbursts which are nearly as bright as the initial one. These recur on a time-scale of approximately 25 days but there is also considerable short-term activity with variations of half a magnitude on a daily basis. The exact shape of the initial outburst is not completely clear as this also show short-term activity. The general behaviour of the eruption is consistent with the `jitter' class of Strope \etal  \cite{2010AJ....140...34S}
and it seems most likely that the lower envelope of the eruption is consistent with a single power-law with $\alpha_1 = 0.9$. A lower limit to the first break point is given by the length of the data, so $t_{b1} > 160$. The values of $t_2 = 111$ and $t_3 = 153$ days are measured from the light-curve and these are also consistent with this class of nova.

There are two possibly contentious issues. The first is that the $CV$ magnitudes are not a good proxy for $V$, but it has been shown that for much of the eruption $CV = V = R$, although the faintest magnitudes may be too bright relative to $V$ by some probably not significant, but unknowable amount. The second issue is that the lower envelope of the eruption has not been sampled and it is certainly true that the statistics are poor, but again this is in the domain of the unknown. Having said that, all the points that appear to define the lower envelope are consistent and provide a result that is not unreasonable.


\begin{thebibliography}{10}
	\newcommand{\enquote}[1]{`#1'}
	
	\bibitem{1995CAS....28.....W}
	B.~{Warner}, \textit{{Cataclysmic Variable Stars}},
	\textit{\href{https://ui.adsabs.harvard.edu/abs/1995CAS....28.....W}{Cambridge
			Astrophysics Series}}, vol.~28 (Cambridge University Press), 1995.
	
	\bibitem{2008clno.book.....B}
	M.~F. {Bode} \& A.~{Evans} (eds.), \textit{{Classical Novae}},
	\textit{\href{https://ui.adsabs.harvard.edu/abs/2008clno.book.....B}{Cambridge
			Astrophysics Series}}, vol.~43 (Cambridge University Press), 2008.
	
	\bibitem{2016PASP..128e1001S}
	S.~{Starrfield}, C.~{Iliadis} \& W.~R. {Hix},
	\textit{\href{https://ui.adsabs.harvard.edu/abs/2016PASP..128e1001S}{\pasp}},
	\textbf{128}, 051001, 2016.
	
	\bibitem{2012ApJ...756..107S}
	M.~M. {Shara} \textit{et~al.},
	\textit{\href{https://ui.adsabs.harvard.edu/abs/2012ApJ...756..107S}{\apj}},
	\textbf{756}, 107, 2012.
	
	\bibitem{2016Natur.537..649M}
	P.~{Mr{\'o}z} \textit{et~al.},
	\textit{\href{https://ui.adsabs.harvard.edu/abs/2016Natur.537..649M}{\nat}},
	\textbf{537}, 649, 2016.
	
	\bibitem{2013IAUS..281..154A}
	G.~C. {Anupama}, in \textit{Binary Paths to Type Ia Supernovae Explosions}
	(R.~{Di Stefano}, M.~{Orio} \& M.~{Moe}, eds.), 2013,
	\textit{\href{https://ui.adsabs.harvard.edu/abs/2013IAUS..281..154A}{IAU
			Symposium}}, vol. 281, pp. 154--161.
	
	\bibitem{2015AcPPP...2..246M}
	K.~{Mukai},
	\textit{\href{https://ui.adsabs.harvard.edu/abs/2015AcPPP...2..246M}{Acta
			Polytechnica CTU Proceedings}}, \textbf{2}, 246, 2015.
	
	\bibitem{MPEnovavM31}
	\enquote{{Optical Novae and Candidates in M 31}}, \\
	\url{http://www.mpe.mpg.de/~m31novae/opt/m31/index.php}, 2020.
	
	\bibitem{kojinovae}
	K.~{Mukai}, \enquote{{List of Recent Galactic Novae}}, \\
	\url{https://asd.gsfc.nasa.gov/Koji.Mukai/novae/novae.html}, 2020.
	
	\bibitem{2014A&A...563L...9D}
	M.~J. {Darnley} \textit{et~al.},
	\textit{\href{https://ui.adsabs.harvard.edu/abs/2014A&A...563L...9D}{\aap}},
	\textbf{563}, L9, 2014.
	
	\bibitem{2017ApJ...849...96D}
	M.~J. {Darnley} \textit{et~al.},
	\textit{\href{https://ui.adsabs.harvard.edu/abs/2017ApJ...849...96D}{\apj}},
	\textbf{849}, 96, 2017.
	
	\bibitem{2017TNSTR.954....1Z}
	S.~{Zhang} \textit{et~al.},
	\textit{\href{https://ui.adsabs.harvard.edu/abs/2017TNSTR.954....1Z}{Transient
			Name Server Discovery Report}}, \textbf{2017-954}, 2017.
	
	\bibitem{2017ATel10741....1W}
	S.~C. {Williams} \& M.~J. {Darnley},
	\textit{\href{http://adsabs.harvard.edu/abs/2017ATel10741....1W}{The
			Astronomer's Telegram}}, \textbf{10741}, 2017.
	
	\bibitem{2017ATel11070....1T}
	H.~{Tan} \textit{et~al.},
	\textit{\href{http://adsabs.harvard.edu/abs/2017ATel11070....1T}{The
			Astronomer's Telegram}}, \textbf{11070}, 2017.
	
	\bibitem{2017ATel11076....1V}
	A.~{Valcheva} \textit{et~al.},
	\textit{\href{http://adsabs.harvard.edu/abs/2017ATel11076....1V}{The
			Astronomer's Telegram}}, \textbf{11076}, 2017.
	
	\bibitem{2016yCat.2336....0H}
	A.~A. {Henden} \textit{et~al.},
	\textit{\href{https://ui.adsabs.harvard.edu/abs/2016yCat.2336....0H}{VizieR
			Online Data Catalog}}, II/336, 2016.
	
	\bibitem{2006AJ....131.2478M}
	P.~{Massey} \textit{et~al.},
	\textit{\href{https://ui.adsabs.harvard.edu/abs/2006AJ....131.2478M}{\aj}},
	\textbf{131}, 2478, 2006.
	
	\bibitem{2014MNRAS.440.3402M}
	U.~{Munari} \textit{et~al.},
	\textit{\href{https://ui.adsabs.harvard.edu/abs/2014MNRAS.440.3402M}{\mnras}},
	\textbf{440}, 3402, 2014.
	
	\bibitem{2015gacv.workE..56C}
	D.~{Chochol} \textit{et~al.}, in
	\textit{\href{https://ui.adsabs.harvard.edu/abs/2015gacv.workE..56C}{The
			Golden Age of Cataclysmic Variables and Related Objects - III (Golden2015)}},
	2015, p.~56.
	
	\bibitem{2015MNRAS.447.1661M}
	U.~{Munari} \textit{et~al.},
	\textit{\href{https://ui.adsabs.harvard.edu/abs/2015MNRAS.447.1661M}{\mnras}},
	\textbf{447}, 1661, 2015.
	
	\bibitem{2016IBVS.6162....1M}
	U.~{Munari} \textit{et~al.},
	\textit{\href{https://ui.adsabs.harvard.edu/abs/2016IBVS.6162....1M}{\ibvs}},
	\textbf{6162}, 1, 2016.
	
	\bibitem{2010AJ....140...34S}
	R.~J. {Strope}, B.~E. {Schaefer} \& A.~A. {Henden}, \textit{\href
		{http://adsabs.harvard.edu/abs/2010AJ....140...34S}{\aj}}, \textbf{140}, 34,
	2010.
	
	\bibitem{2006ApJS..167...59H}
	I.~{Hachisu} \& M.~{Kato},
	\textit{\href{https://ui.adsabs.harvard.edu/abs/2006ApJS..167...59H}{\apjs}},
	\textbf{167}, 59, 2006.
	
\end{thebibliography}

\OBSversion 
\end{document}